\documentclass[runningheads]{llncs}
\usepackage[T1]{fontenc}
\usepackage{graphicx} 
\usepackage{enumitem}
\usepackage{wrapfig}

\usepackage{cite}
\title{Asynchronous Checkpoint for Eventually Consistent Databases
}
\author{Raaghav Ravishankar \inst{1}
\and Sandeep Kulkarni \inst{1}
\and Nitin H. Vaidya \inst{2}
}
\institute{Michigan State University \and Georgetown University }

\usepackage{xspace}
\usepackage{tabularx}
\usepackage{amssymb,amsmath}
\usepackage{mathtools}

\usepackage{subcaption}
\usepackage{color}
\usepackage[plain, linesnumbered, noend, noresetcount]{algorithm2e}
\makeatletter
\renewcommand{\@algocf@capt@plain}{above}
\renewcommand{\algocf@caption@plain}{\box\algocf@capbox\vskip\AlCapSkip}%
\makeatother
\usepackage{comment}
\usepackage{cleveref}
\usepackage{pdflscape}
\usepackage{url}
\usepackage{multicol}
\usepackage{lscape}
\crefname{algocf}{alg.}{algs.}
\Crefname{algocf}{Algorithm}{Algorithms}
\SetKwProg{Fn}{}{:}{end}
\SetKwProg{Struct}{struct}{ \{}{\}\;}
\SetKwComment{Comment}{\slash \slash }{}

\definecolor{asparagus}{rgb}{0.13, 0.54, 0.13}
\newcommand{\ifPaperMode}{\iffalse}

\newcommand{\rnote}[1]{{\color{blue}#1}}
\newcommand{\snote}[1]{{\color{asparagus}#1}}
\newcommand{\bnote}[1]{{\color{red}#1}}

\newcommand{\nitin}[1]{{\color{red}\bf NV: #1}}

\newcommand{\commitLog}{\tt{commit\mbox{--}log}\xspace}

\newcommand{\MessageSender}{\tt{MessageSender}\xspace}
\newcommand{\MessageReceiver}{\tt{MessageReceiver}\xspace}
\newcommand{\CutForCheckpoint}{\tt{CPRequest}\xspace}
\newcommand{\CutForCheckpointReply}{\tt{CPReply}\xspace}
\newcommand{\CheckpointCutEntry}{\tt{CpLog}\xspace}
\newcommand{\AsyncCheckpointer}{\tt{AsyncCheckpointer}\xspace}
\newcommand{\ExecuteTransaction}{\tt{ExecuteTxn}\xspace}
\newcommand{\MessageCounter}{\tt{MsgCount}\xspace}
\newcommand{\GreenMessageCounter}{\tt{GreenCounts}\xspace}
\newcommand{\RedMessageCounter}{\tt{RedCounts}\xspace}

\newcommand{\ReplicaColor}{\tt{RColor}\xspace}
\newcommand{\currentColor}{\tt{currColor}\xspace}
\newcommand{\hb}{\ \ensuremath{{\mathcal{\textbf{hb}}}}\ }

\newcommand{\DTCC}{DTCS\xspace}

\newcommand{\MuFASA}{MuFASA\xspace}

\usepackage{nccmath}

\newtheorem{observation}{Observation}

\begin{document}

\maketitle



\begin{abstract}

We focus on the problem of checkpointing (or taking a snapshot) in fully replicated eventually consistent distributed databases. In particular, we consider the problem of taking Distributed Transaction-Consistent Snapshots (\DTCC). A typical example of such a system is a replicated main-memory database that provides strong eventual consistency. This problem is important and challenging for several reasons: (1) eventual consistency often creates anomalies that the users do not anticipate. Hence, frequent snapshots that can be used to ascertain desired invariants are highly beneficial in their maintenance, and (2) traditional distributed snapshot algorithms lead to significant overhead and/or inconsistencies such as storing dirty writes of incomplete transactions.

A key benefit of \DTCC is that it summarizes the computation by a sequence of snapshots that are strongly consistent even though the underlying computation is only weakly consistent.
In essence, when anomalies arise in an eventually consistent system, \DTCC enables one to concentrate solely on the snapshots surrounding the time point of the anomaly.

By showing that traditional distributed snapshots lead to inconsistencies and/or excessive overhead, we define the notion of size-minimal \DTCC for fully replicated databases. We present \MuFASA, an algorithm for a size-minimal \DTCC with minimal checkpointing overhead (only $O(n)$ new messages and the addition of a single counter for existing messages). \MuFASA also provides a significant benefit over existing checkpointing algorithms for distributed systems and replicated main-memory databases by being a fully asynchronous protocol.

\end{abstract}


\section{Introduction} \label{sec:Introduction}

Consider replicated database systems such as \cite{riak,redis,Dynamo,voldemortdb,AntidoteDB}, which provides eventual consistency. In an eventually consistent system, in the absence of new updates, all replicas will eventually converge to an identical state. However, in general, transactions in such a system may observe inconsistent data. Eventual consistency is sometimes favored in practice for its performance advantages, particularly in distributed environments where high availability is critical.

One challenge of eventually consistent databases is the inconsistency\cite{provingcorrectnessoftransformation} that a user may observe. One way to minimize the effect of this inconsistency is to periodically check desired invariant properties. This can be obtained by stopping the system and allowing it to converge. The goal of this paper is to provide this ability in a \textbf{fully asynchronous manner}.

\subsection{System Model}    



    \begin{wrapfigure}{r}{0.62\textwidth}
        \vspace{-0.8cm}
        \includegraphics[width=0.6\textwidth]{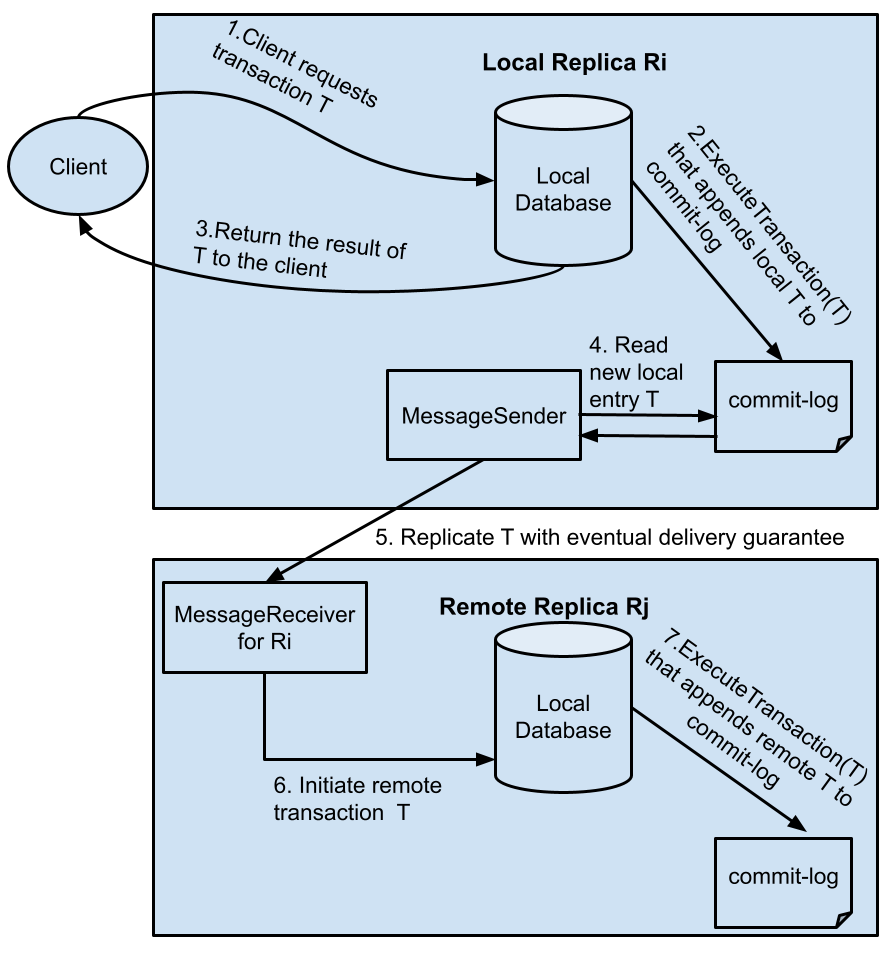}
        
        \caption{Transaction execution}
        
        \label{fig:ModelofDatabase}

        \vspace{-0.5cm}
    \end{wrapfigure}

The system consists of $n$ database replicas named $R1$, $R2$,$\cdots$, $Rn$. The replicas communicate via pairwise unidirectional non-FIFO channels. Each channel eventually delivers each message sent on that channel. For any control messages sent by our algorithm, separate non-FIFO directed channels are used between each pair of replicas.

The database consists of a set of objects. Each replica has a copy of every object in the database.
Each transaction may read or write multiple objects.
A client requests one of the replicas to perform its transaction. As shown in step 1 in \Cref{fig:ModelofDatabase}, suppose that a client requests replica $Ri$ to execute a transaction T. We will say that transaction T was ``issued at'' $Ri$.



Replica $Ri$ then performs transaction T. Once T begins, we say it leaves $Ri$ in an \textit{uncommitted} (i.e., dirty or inconsistent) state until it finishes. When T commits, it updates its $\commitLog$, for durability (step 2 in the figure). We refer to this moment as the commit time of T in $R_i$. Standard concurrency control by two-phase locking \cite{databasesystemconceptsbook} is employed within each replica so that the transactions performed at that individual replica achieve the strict serializability property (i.e., the transactions issued at each individual replica are totally ordered in the order of their commit times at that replica).
%
Our assumptions are similar to those made in other related work \cite{databasecheckpointsurvey,CALC,lowcostdatabaseCP2001}.

A $\MessageSender$ thread reads the $\commitLog$ at replica $Ri$ for issued transactions in the log, and eventually sends an ``update'' message to the other replica. This update message will specify the updates performed by committed transaction T (steps 4 and 5).

A $\MessageReceiver$ thread at each of the other replicas (say  $Rj$) will eventually receive the update message from replica $Ri$, and make corresponding updates to the state of the local copy of the objects (i.e., the objects in $Rj$). (step 6)
In particular, on receipt of the update message corresponding to transaction T from its issued replica $Ri$, the $\MessageReceiver$ at replica $Rj$ initiates transaction T at $Rj$, which performs all the updates specified for transaction T in the update message; when this transactions T is completed at $Rj$, the $\commitLog$ at replica $Rj$ is updated to reflect its completion (steps 6 and 7). Note that, at replica $Rj$, transaction T is serialized with respect to the transactions initiated at replica $Rj$. However, it is important to note that, the order in which transactions issued by different replicas are serialized at different replicas may be different. 

When performing such updates at replica $Rj$, any write-conflicts are resolved using standard conflict resolution mechanisms that have been proposed for eventually consistent systems (such as the ``last writer wins'' rule \cite{PutConsistencyInEventual}, or commutative functions of CmRDTs \cite{CRDTs}). The use of such a conflict resolution mechanism is to ensure that the system achieves the eventual consistency property.

In the above example, observe that transaction $T$ is ``issued at'' only one replica, namely $Ri$. In this case, transaction $T$ is initially performed at replica $Ri$. When another replica, such as $Rj$, eventually receives the update corresponding to transaction $T$,  transaction $T$ is also ``performed at'' replica $Rj$. The order in which different transactions are performed at different replicas can be different. The ordering of any transaction is imposed by the strict serialization at the issued replica. If there are two conflicting, parallel transactions issued by clients to different replicas, they might execute in a different order at each replica but the final state becomes identical due to the conflict resolution mechanisms to maintain eventual consistency.


\subsection{Distributed Transaction-Consistent Snapshots (DTCS)}
\label{subsec:DTCS}
Our goal in this paper is to develop an algorithm that records a ``distributed transaction-consistent snapshot (DTCS)" of the replicated database system. 
%
%
We now list the properties that a DTCS must satisfy:

\begin{itemize}
\item \textbf{Dirty Write Exclusion:} A DTCS corresponds to the state of the database after the execution of a subset of {\it committed} transactions (no uncommitted/dirty writes included). Let us consider a DTCS named $CP$ (i.e., checkpoint). Then the set of committed transactions corresponding to $CP$ is denoted as $CP_{set}$.
\item {\bf Dependency Closure Property:} Suppose that transaction $T\in CP_{set}$. Suppose that transaction $T$ was committed at its issued replica $Ri$ after some transaction $U$ (issued at any replica) was committed at $Ri$, then $U$ must be included in $CP_{set}$.

\end{itemize}






Since the transactions issued at each replica are totally ordered, the set of transactions included in a DTCS includes a prefix of the transactions committed at each replica.

%

\begin{figure}[tbh]
\begin{subfigure}{0.5\linewidth}
    \includegraphics[height=3cm]{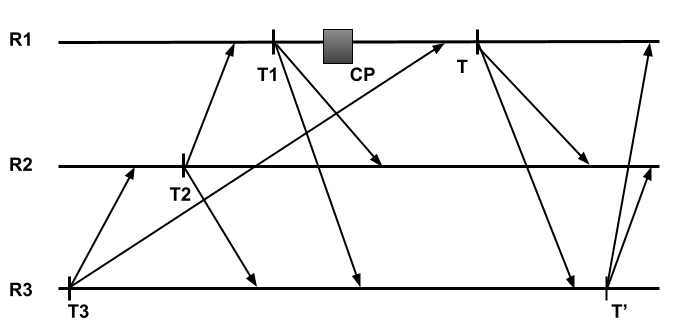}
    \caption{\DTCC needs to ensure T3 is included and T' is excluded in the final checkpoint.}
    \label{fig:DifficultyOfDTCS}
\end{subfigure}
\hfill
\begin{subfigure}{0.5\linewidth}
    \includegraphics[height=3cm]{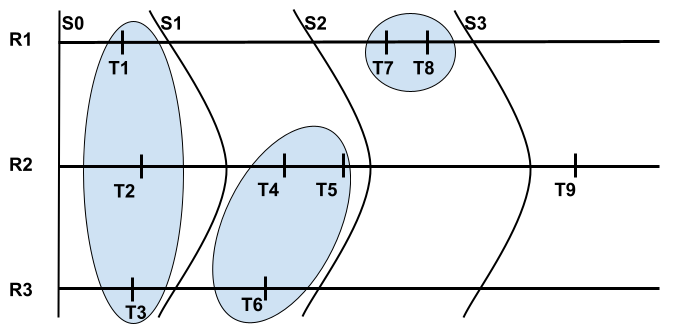}
    \caption{Sequence of checkpoints}
    \label{fig:supertransaction}
\end{subfigure}

\caption{Visualization of DTCS. Transaction times in the timelines are the respective commit times at that replica.}
\end{figure}

Note that we also refer to a snapshot as a ``checkpoint'' in our discussion.
Consider the following example, as illustrated in Figure \ref{fig:DifficultyOfDTCS} (marked transaction times are their commit times at the respective replica).
\begin{itemize}[wide, labelwidth=!, labelindent=0pt, topsep=0pt]
\item Transactions T1, T2 and T3 are issued at replicas $R1$, $R2$ and $R3$, respectively.
\item Transaction T3 commits at replica $R3$, and update messages for T3 are sent to replicas $R1$ and $R2$.
\item Replica $R2$ receives the above update message from replica $R3$; replica R2 then performs transaction T3 locally and commits T3. Then replica $R2$ commits transaction T2, and sends updates for T2 to replicas $R1$ and $R3$.
\item Replica $R1$ now receives the update for T2 from $R2$, and performs T2 locally.  Replica $R1$ then commits transaction T1.
\end{itemize}

\vspace{-0.6cm}
\vspace{-0.3cm}

We make a few observations:
\begin{itemize}
\item Due to the dependencies introduced through the update messages, if T1 is included in $CP_{set}$ then T2 and T3 must also be included in $CP_{set}$.

\item Replica $R2$ commits T3 before committing T2, whereas $R1$ commits T2 before committing T3 (because $R1$ receives the update for T3 from $R3$ only after $R1$ has already committed T2). Despite this difference in the commit order, the final state of the replicas would be identical, due to the use of the conflict-resolution mechanisms used for implementing eventual consistency.

\item Suppose that the snapshot at $R1$ ended at CP. Transaction T issued at $R1$ after CP is not included in the checkpoint. This implies that transaction T' should not be a part of the  checkpoint (including T' while not including T will violate the dependency closure property).

\item Because the transactions issued at any given replica are serialized at that replica, in the above example, all locally issued transactions committed at $R1$ before T1 must be in $CP_{set}$ (the figure does not illustrate any such transactions, however). Similarly, all locally issued transactions committed at $R3$ before T3 must also be in $CP_{set}$.

\end{itemize}

\noindent{\bf Repeated invocations of DTCS:}
Repeated non-overlapping invocations of \DTCC (each following the completion of the previous one) will yield an execution trace resembling Figure~\ref{fig:supertransaction}. We refer to the snapshot boundaries $S1$, $S2$, $S3$ as \textit{global cuts}\cite{vijaygargscalableCP}, and their intercepts with each replica (at the local times of the replica) as \textit{local cuts}. In this trace, $CP_{set}^i$ represents the set of all transactions captured in the $i^{th}$ snapshot, and $Si$ denotes the corresponding system state. We can observe the following: (1) $S1$ can be obtained by executing a ``super-transaction'' composed of all transactions in $CP_{set}^1$ from the initial state, i.e., $S0$, (2) $S2$ can be obtained by executing a super-transaction consisting of all transactions in $(CP_{set}^2\!-\!CP_{set}^1)$ starting from $S1$ (or by executing $CP_{set}^2$ from $S0$), and so on. 
Thus, when viewed as a sequence of super-transactions, \DTCC ensures its serializable execution. 
Thus, when viewed as a sequence of super-transactions, execution in Figure \ref{fig:supertransaction} is intuitively a ``serializable execution'' where super-transaction $S1$ (consisting of T1, T2, T3) occurs before $S2$ (consisting of T4, T5, T6) which occurs before $S3$ (consisting of T7 and T8). 
In other words, \textbf{\DTCC provides a ``serializable trajectory'' within the broader landscape of eventual consistency}.




~\\
As stated earlier, we propose an algorithm for taking distributed transaction-consistent snapshots (DTCS). We now list the properties we desire from the DTCS algorithm.

    \begin{enumerate}[wide, labelwidth=!, labelindent=0pt, topsep=0pt]
        \item {\bf Recency}: This is a progress property. We will designate a replica as an initiator of the \DTCC algorithm. We require that all transactions that have been committed in the initiator replica at the time of the invocation of the DTCS algorithm be reflected in the DTCS (i.e., the checkpoint) recorded by the  algorithm.
      
        \item \textbf{Strict Concision:} The information stored as part of the checkpoint should include only one copy of each object in the database. Since the database may contain a large number of objects, this property would help minimize the overhead of storing a checkpoint.
        This also allows one to verify desired invariant properties easily if the entire snapshot is available on one replica. It also simplifies rollback in the future.

        \item \textbf{Asynchronous behavior:} The \DTCC algorithm should not quiesce the distributed database, and the normal operation of the database should proceed in parallel with the DTCS algorithm. In particular, the control messages sent by the DTCS algorithm should not block the update messages sent by the replicas.
        The asynchronous behavior is intended to reduce performance degradation for the transactions despite the use of a DTCS algorithm.


        \item \textbf{Bounded message complexity:} The total number of additional messages sent by the \DTCC algorithm should not depend on the number of objects in the database or the number of transactions that have taken place in the system.
        
    \end{enumerate}

\subsection{A Simple Algorithm and Its Shortcomings}
\label{subsec:chandylamportbad}
Chandy and Lamport \cite{chandylamport} proposed an elegant algorithm for taking a snapshot of a distributed system wherein the processes use message-passing for communication. Since our distributed database is layered on top of a message-passing system, the Chandy-Lamport algorithm can be applied in our setting too.

In brief, in the Chandy-Lamport algorithm, an ``initiator'' begins a checkpoint (or distributed snapshot) by atomically recording its own local state and sending a ``marker'' message on all its outgoing channels. When any other process receives a marker for the first time, it atomically records its own state and sends markers on all of its outgoing channels. The algorithm also specifies which messages should be recorded as part of the snapshot.

The above algorithm is designed to work with FIFO channels, however, variants of this algorithm have been proposed that also work with non-FIFO channels \cite{LaiYangGlobalSnapshot, MATTERN1993423,ajayCP,vijaygargscalableCP}. The Chandy-Lamport algorithm, and its variants, have two  shortcomings in our context:
\begin{itemize}
    \item Each process needs to record its local state {\it atomically}. In the context of our problem, when a replica is recording its local state, no transactions would be allowed to commit. This blocking behavior conflicts with the expectation for asynchronous behavior from the DTCS algorithm.

    \item In Chandy-Lamport algorithm, each process records its own local state. When applied to our system, each replica would record its state, which contains one copy of each object in the database. Thus, collectively, the replicas would save $n$ copies of each object (where $n$ is the number of replicas).

    \item When used without any modification, the captured local states might contain dirty writes from transactions that are still uncommitted by that time point.
\end{itemize}


We address these challenges using an initiator replica to capture an initial snapshot. A novel asynchronous three-color system then updates the remaining distributed state at the initiator itself while preserving \DTCC properties.

\subsection{Contributions and Paper Outline}

Our goal in this paper is to present an algorithm that has all the desirable properties listed above. We present \MuFASA, a \underline{Mu}lti-replica \underline{F}ully \underline{A}synchronous \underline{S}napshot \underline{A}lgorithm. This algorithm guarantees that the snapshot is size-minimal and contains exactly one copy of each object in the database. It ensures that it is fully asynchronous, i.e., it does not block the underlying database transactions in any manner. Its overhead is minimal, needing only $O(n)$ control messages, where $n$ is the number of replicas (as compared to $O(n^2)$ in many distributed snapshot algorithms). 


The rest of this paper is organized as follows. \Cref{sec:Background} discusses the related work.
\Cref{sec:BriefAlgorithm} briefly describes our algorithm \MuFASA and how it achieves the desired properties listed above. 
Due to lack of space, a detailed description of the algorithm and correctness arguments are presented in the appendix.
Finally, \Cref{sec:Conclusion} summarizes the paper and provides directions for future work.
\section{Related Work}\label{sec:Background} 

We first discuss the related work on checkpointing in non-transaction oriented systems, followed by the related work on checkpointing in transactional database systems. \Cref{tab:ComparisonOfAlgorithms} compares our contributions with relevant past work.

\subsection*{Checkpointing in Distributed Systems}

        Chandy and Lamport\cite{chandylamport} introduced the notion of a global snapshot. For a $n$-process system, it uses $O(n^2)$ control messages and also requires FIFO channels.  
        Lai and Yang \cite{LaiYangGlobalSnapshot}  illustrated a global snapshotting algorithm that allowed non-FIFO channels. 
        However, it required a significant overhead to exchange the set of all sent/received messages to identify the channel state.
        
        Mattern \cite{MATTERN1993423} improved the algorithm of Lai and Yang by using counters that obviated the need to exchange message sets. 
        This still requires $O(n^2)$ control messages, but each message is an integer. 
        
        Garg et al. \cite{vijaygargscalableCP} improved on Mattern \cite{MATTERN1993423} by observing that the technique can be optimized by looking at checkpointing as a distributed counting problem. With this, the number of control messages exchanged was reduced to $O(n\log w)$, where $w$ is the average number of messages in each channel. Kshemkalyani \cite{ajayCP} introduced snapshots for networks with a hypercube topology (unlike a fully connected topology presented in this paper) with $O(n)$ message complexity and $O(n^2)$ piggybacking on messages.  

    \subsection*{Checkpointing in Database Systems}

        The notion of a transaction-consistent checkpoint was explored first by DeWitt et al. \cite{PrinciplesOfTCC}, where a checkpoint is transaction-consistent if it only contains the effect of all committed transactions but none of the ongoing uncommitted (dirty) transactions. A significant number of industrial methods \cite{databasecheckpointsurvey} utilize the fork() operating system call(\cite{fork, MPIFork}) to create a child process state to capture the checkpoint.
        However, this approach causes a latency spike proportional to the size of the database and gives the appearance that the database is $quiesced$. Several asynchronous checkpointing algorithms have been studied to overcome this latency spike. 
            As state of the art, Cao et al. introduced Zig-Zag and Ping-Pong\cite{frequentlyconsistent} checkpointing algorithms for applications that are frequently consistent, where the application triggers a checkpointer whenever it becomes consistent. However, in databases with a very high unimpeded update rate, these approaches would not be applicable.

            To overcome the limitation in \cite{frequentlyconsistent}, Ren et al. introduced CALC\cite{CALC}, which does not require the application to reach such a physical point of consistency. Although it also performs copy-on-write (as in \cite{frequentlyconsistent}) whenever it desires to take a checkpoint, it also uses shared memory techniques to wait until a point of consistency is simulated. This simulated point is referred to as a virtual point of consistency that generates a view of the database that includes only committed transactions up to that point, and avoids the dirty effects of uncommitted transactions that are happening at that point in time. 

            Li et al. studied the Redis database with all the state-of-the-art approaches in checkpointing, and provided improved techniques for the Redis database, called HourGlass and Piggyback\cite{databasecheckpointsurvey}. Much like CALC, the approach avoids latency spikes proportional to the size of the database, and the approach conclusively scales with the database size better than the OS fork() method.

            Lin and Dunham \cite{lowcostdatabaseCP2001} explored checkpointing in partitioned (distributed) databases by taking a collection of \textit{fuzzy} checkpoints and additional logs to \textit{undo} the uncommitted transactions that would otherwise be part of the fuzzy checkpoint. The fuzzy checkpoints thus generated utilize loosely synchronized techniques unlike the asynchronous approach in this paper.
            Using their approach in a replicated database will result in massive data redundancy, as many copies of the object will be stored in the checkpoint. 

    An important difference between a checkpointing problem for generic distributed systems and our checkpointing goals for a fully replicated database is that we not only require a consistent global snapshot, but we also require that - {(1) the size of the stored checkpoint is minimized by avoiding redundancy, and (2) the snapshot does not include the dirty effects of uncommitted transactions. This is done by combining the state of all replicas into one transaction consistent view of the replicated database.

\begin{table}[!h]
    \vspace{-0.5cm}
    \centering
    \begin{tabularx}{\textwidth}{|>{\centering\arraybackslash\hsize=1.6\hsize\linewidth=\hsize}X|
    >{\centering\arraybackslash\hsize=1\hsize\linewidth=\hsize}X|
    >{\centering\arraybackslash\hsize=.55\hsize\linewidth=\hsize}X|
    >{\centering\arraybackslash\hsize=.55\hsize\linewidth=\hsize}X|
    >{\centering\arraybackslash\hsize=.55\hsize\linewidth=\hsize}X|
    >{\centering\arraybackslash\hsize=.75\hsize\linewidth=\hsize}X|
    >{\centering\arraybackslash\hsize=.75\hsize\linewidth=\hsize}X|
    >{\centering\arraybackslash\hsize=.55\hsize\linewidth=\hsize}X|}
        \hline
         Attribute $\rightarrow$ Algorithm $\downarrow$ & Distributed Snapshot & TCC & Async. & Size Minimal & Control Messages & Non-FIFO channels & Instant Rollback \\
         \hline
         Chandy \& Lamport\cite{chandylamport} & $\checkmark$ &  &  & $\checkmark$ & O($n^2$) & & \\
         \hline
         Lai \& Yang\cite{LaiYangGlobalSnapshot} & $\checkmark$ &  &  & $\checkmark$ & O($n^2$) & $\checkmark$ & \\
         \hline
         Mattern\cite{MATTERN1993423} & $\checkmark$ &  &  & $\checkmark$  & O($n^2$) & $\checkmark$ & \\
         \hline
         Garg et.al\cite{vijaygargscalableCP} & $\checkmark$ &  &  & $\checkmark$  & O(nlogw) &  $\checkmark$  & \\
         \hline
         Cao et al.\cite{frequentlyconsistent} &  & $\checkmark$  &  & $\checkmark$ & N/A & N/A & $\checkmark$ \\
         \hline
         Ren et al.\cite{CALC} &  & $\checkmark$  & $\checkmark$  & $\checkmark$ & N/A & N/A & $\checkmark$ \\
         \hline
         Li et al.\cite{databasecheckpointsurvey} &  & $\checkmark$  & $\checkmark$  &  & N/A & N/A & $\checkmark$ \\
         \hline
         Lin \& Dunham\cite{lowcostdatabaseCP2001} & $\checkmark$ & $\checkmark$ & & & O(n) & $\checkmark$ & \\
         \hline
         \MuFASA & $\checkmark$ & $\checkmark$ & $\checkmark$ & $\checkmark$ & O(n) & $\checkmark$ & $\checkmark$  \\
         \hline

    \end{tabularx}
    \caption{Comparison with existing algorithms when used for replicated databases}
            \label{tab:ComparisonOfAlgorithms}
\vspace{-1cm}
\end{table}

\section{\underline{Mu}lti-replica \underline{F}ully \underline{A}synchronous \underline{S}napshot \underline{A}lgorithm }
\label{sec:BriefAlgorithm}

Before sketching the proposed \MuFASA algorithm, we list some of its features:
\begin{itemize}[wide, labelwidth=!, labelindent=0pt, topsep=0pt]
    \item One of the replicas is designated as the initiator of the algorithm. In our presentation below, we assume that replica $R1$ is the initiator. In general, the initiator can be changed over time by the agreement of the replicas in the system (for instance, using a Round-Robin policy).
    \item As the initiator, only replica $R1$ saves a copy of each object in the database as a part of the checkpoint. The other replicas help replica $R1$ in determining which object versions should be saved as part of the checkpoint. This helps us minimize the storage overhead of the checkpoint, by avoiding redundant copies of the objects from being saved by different replicas.
    \item Separate communication channels are used for control messages required for implementing the \MuFASA algorithm. In particular, the control messages do not block delivery of the update messages exchanged by the database mechanisms.

    \item Replica colors and message colors:
    \begin{itemize}
        \item At any point of time, each replica is assigned one of the following colors: Green, Yellow or Red. The color is maintained in each replica using a shared memory variable $\ReplicaColor$, and the accesses to this variable at each replica are linearizable. Note that multiple threads at each replica may access the $\ReplicaColor$ variable at that replica. We assume that compare-and-swap operation may be performed on $\ReplicaColor$. We use this operation to ensure that updates to $\ReplicaColor$ occur only from Green to Yellow or Red, and from Yellow to Red. This is useful because multiple $\MessageReceiver$ threads might want to update the color concurrently, as seen later.
        \item Each message sent by each replica also contains two bits to indicate the message color. Each message sent by a replica has the same color as the color of that replica at the time of sending that message, with the following exception to this rule: When replica $R1$ is Yellow, the messages it sends are Green. Thus, replica $R1$ does not send Yellow-colored messages. The $\MessageSender$ of each replica reads the $\ReplicaColor$ variable before sending each message to decide the color to be assigned to that message.\\
    \end{itemize} 

\end{itemize}

As alluded above, the steps performed by replica $R1$ are different from the other replicas. We will describe the steps for $R1$ and other replicas separately below.
We consider only a single invocation of the algorithm. When multiple invocations are considered, the invocation number would be added to the state variables (such as the colors) appropriately. For brevity, we exclude these details.



\subsection*{Operations at the Initiator Replica $R1$}
\label{subsec:InitiatorOperations}
Replica $R1$ has Green color initially.
    When $R1$ initiates a new instance of the DTCS algorithm to take a transaction-consistent checkpoint of the database, it first records its own state. $R1$ records its own state using an asynchronous algorithm that does not require blocking of the execution of the transactions at $R1$. In particular, we use the CALC algorithm \cite{CALC} for this purpose. Rather than describing CALC fully, we summarize the properties that it achieves:
    \begin{itemize}
        \item CALC records the state of $R1$ as a result of a certain prefix of transactions committed at $R1$, with the additional properties below.
        \item The effects of all transactions that were committed in $R1$ prior to the start of CALC are included in the local checkpoint of $R1$. Note that when these transactions were initiated, $\ReplicaColor$ at $R1$ would have been Green. 
        
        \item When replica $R1$ initiates CALC, its $\ReplicaColor$ is changed to Yellow. The effects of the transactions that were initiated at $R1$ when $R1$ had Yellow color may or may not be included in the local checkpoint of $R1$.
        (Recall that replica $R1$'s messages are colored Green when $R1$'s color is either Yellow or Green.)
        
        \item When $R1$ completes CALC, it would have recorded a checkpoint of its local state (with the properties above), and its color is changed to Red \footnote{Changing of $R1$ color from Yellow to Red is analogous with the CALC algorithm\cite{CALC} changing its phase variable from PREPARE to RESOLVE.}. As may be expected, the effects of the transactions that are initiated at $R1$ when $R1$ is Red are excluded from the local checkpoint of $R1$, as recorded by CALC. After this point, $\MessageSender$ at $R1$ will color its messages in Red.
    \end{itemize}
   \Cref{tab:TransactionColoringAtR1} summarizes effort of each transaction that commits at $R1$.To implement these color changes, we make minor modifications to the steps in the CALC algorithm performed by $R1$ (these modifications are described in the appendix).
   In brief, CALC maintains a backup copy of each object before updating it to a new value in its transactions by utilizing copy-on-write techniques. The CALC checkpointer collects the final values of the backup copy of each object for its checkpoint. As shown in \Cref{tab:TransactionColoringAtR1}, if a transaction should include itself in the checkpoint, it updates the backup copy with its current update value. Otherwise, it does not update the backup copy to exclude the transaction.

   The operations performed by the non-initiator replicas, and how replica $R1$ interprets them, is described next. 

\begin{table}[!h]
            \centering
            \begin{tabularx}{\textwidth}{|>{\centering\arraybackslash\hsize=.35\hsize\linewidth=\hsize}X|
            >{\centering\arraybackslash\hsize=.55\hsize\linewidth=\hsize}X|
            >{\centering\arraybackslash\hsize=.55\hsize\linewidth=\hsize}X|
            >{\centering\arraybackslash\hsize=.55\hsize\linewidth=\hsize}X|}
                \hline
                 EndColor $\rightarrow$ StartColor $\downarrow$ & Green & Yellow & Red  \\
                 \hline
                 Green & Automatically included in CP & Automatically included in CP &  Not possible \\
                 \hline
                 Yellow & Not Possible & T takes additional steps to change the CP to include itself & T takes additional steps to exclude itself in the CP \\
                 \hline
                 Red & Not Possible & Not Possible & T takes additional steps to exclude itself from the CP \\
                 \hline
            \end{tabularx}
            \caption{
            Any transaction T in $R1$ checks $\ReplicaColor$ at the start and end of the transaction to decide its inclusion in the checkpoint $CP$.}
            \label{tab:TransactionColoringAtR1}
        \vspace{-1cm}

        \end{table}

\subsection*{Operations at the Non-Initiator Replicas}
\label{subsec:modifiedApproach}

Each non-initiator replica $Ri ~(i \neq 1)$ is initialized with $\ReplicaColor$ Green and the messages it sends are also colored Green. In response to messages received of different colors, $Ri$ takes the following steps
(also summarized in \Cref{tab:ReplicaColor at $R_1$})
:
\begin{itemize}
    \item When a $\MessageReceiver$ of $Ri$ first receives a non-Green (i.e., Yellow or Red) message from any replica, it changes  $\ReplicaColor$ of $Ri$ to Yellow and logs a $\CheckpointCutEntry
    $ entry into the $\commitLog$ before executing updates from the non-Green message. It takes no additional action for the Green messages before this and for subsequent messages of any color.
    \item The $\MessageSender$, upon reading $\ReplicaColor$ before sending updates for the next $\commitLog$ entry, will also log $\CheckpointCutEntry$ into the $\commitLog$ if it sees it to be Yellow for the first time, and then colors its future messages in Yellow.
    \item When the $\MessageSender$ first reads a $\CheckpointCutEntry$ from the $\commitLog$, it changes $\ReplicaColor$ to Red, and colors its future messages in Red.
\end{itemize}

    A replica $Ri ~(i\neq 1)$ performs its local cut when it first appends $\CheckpointCutEntry$ into the $\commitLog$. Because multiple $\MessageReceiver$ threads may receive a non-Green message simultaneously, we use compare-and-swap to set the $\ReplicaColor$ to Yellow only if it still Green. This avoids the problem of writing $\ReplicaColor$ to Yellow after it turns Red. Note that, even after a successful compare-and-swap, the appending of $\CheckpointCutEntry$ by the $\MessageReceiver$ can get pre-empted (race condition between threads). Thus, the $\MessageSender$ also appends this entry. This way, upon receiving the first non-Green message at $Ri$, both the $\ReplicaColor$ is set to Yellow and a $\CheckpointCutEntry$ is appended into the $\commitLog$ to demarcate its local cut before delivering the update message. This asynchronously emulates the global snapshot requirement of performing a cut before processing the first marker message in Chandy and Lamport\cite{chandylamport}.
    
    After $Ri$ turns Yellow, $\MessageSender$ eventually catches up to the $\CheckpointCutEntry$ entry and sets $\ReplicaColor$ to Red, to send future messages in Red. This indicates to the initiator that all updates present after $\CheckpointCutEntry$ are not part of the local cut at $Ri$. $R1$ therefore applies all non-Red received updates from other replicas on top of its local checkpoint, to achieve the final global checkpoint, which is a DTCS.

\begin{table}[!h]

            \begin{tabularx}{\textwidth} {|>{\centering\arraybackslash\hsize=.50\hsize\linewidth=\hsize}X|>{\centering\arraybackslash\hsize=.65\hsize\linewidth=\hsize}X|>{\centering\arraybackslash\hsize=.65\hsize\linewidth=\hsize}X|>{\centering\arraybackslash\hsize=1.4\hsize\linewidth=\hsize}X|>{\centering\arraybackslash\hsize=1.4\hsize\linewidth=\hsize}X|>{\centering\arraybackslash\hsize=1.4\hsize\linewidth=\hsize}X|}
                \hline
                 $Ri$ & $Ri$ Color & Message Color & Significance to $R1$ & Significance to $Ri$ & Significance to $Rj, j\notin \{1,i\}$ \\
                 \hline
                 $R1$ & Green & Green & Checkpoint has not begun & Same because $Ri$ is $R1$ &  Checkpoint has not begun \\
                \hline
                $R1$ & Yellow & Green & Checkpoint has begun & Same because $Ri$ is $R1$ & No Checkpoint action to be taken yet \\
                \hline
                $R1$ & Red & Red & Local Checkpoint Completed & Same because $Ri$ is $R1$ & Local Checkpoint Cut needs to be taken \\
                \hline
                $Ri$, $i\neq 1$ & Green & Green & Transactions are included in Checkpoint & Checkpoint has not begun & Checkpoint has not begun\\
                \hline
                $Ri$, $i \neq 1$ & Yellow & Yellow & Update Checkpoint to include Transactions& Do a local cut before sending messages & Local Checkpoint Cut needs to be taken \\
                \hline
                $Ri$, $i\neq 1$ & Red & Red & Exclude these Transactions from Checkpoint & All included transactions for the local cut have been broadcasted & Local Checkpoint Cut needs to be taken \\ 
                \hline
            \end{tabularx}
            \caption{Color of a Replica $Ri$, its sent message colors and the significance of these colors to every replica.}
            \label{tab:ReplicaColor at $R_1$}
            \vspace{-0.6cm}
        \end{table}

\subsection*{Termination of \MuFASA}
\label{subsec:Termination}
    While the color change will help identify the cut, \MuFASA needs to terminate without needing an update sent along the replication channels. For this purpose, all non-initiator replicas maintain a counter for the number of Green and Yellow messages that they send. $R1$ counts the number of Green and Yellow messages received from each replica in separate counters. After $R1$ turns Red, it broadcasts a $\CutForCheckpoint$ control message to all replicas in a separate control channel to avoid congestion for update messages. When a non-initiator replica $Ri$ receives this message, it treats it as a Red message as described earlier and ensures that $Ri$ turns Yellow (if not already). Later when $Ri$ turns Red, it sends its counter value to $R1$ through the control messages channel. Once $R1$ receives this counter from all replicas, it waits until its own received counters of the Green/Yellow messages match the counters sent by each replica individually. When they match, the initiator knows that all updates up to the desired global cut are now included in the DTCS snapshot and terminates. This efficient termination technique is inspired from Mattern's global snapshot\cite{MATTERN1993423}. Note that in \MuFASA, however, we only send $n-1$ $\CutForCheckpoint$ messages and receive 1 reply for each of them, bounding the message complexity of \MuFASA to $2n-2$ control messages.
      ~\\  
      \vspace{-0.1cm}

      \noindent{\bf Rollback/Recovery:} For crash recovery, all replicas start first from the last stored checkpoint. Then they collect all subsequent issued transactions present in $\commitLog$ (after $\CheckpointCutEntry$ entry) into one list of transactions to be executed by every replica. A replica resumes computation after all transactions in the combined list are committed on top of the checkpointed database. In this way, each replica will gain the effect of all transactions that have taken place up to the point of failure. 


~\\
\noindent{\bf Discussion:}
Here we explain some interesting properties of \MuFASA:
\begin{enumerate} [wide, labelwidth=!, labelindent=0pt, topsep=0pt]
    \item \textbf{Benefits of Bounded Message Complexity and Asynchrony:} 
    \MuFASA utilized only $2n-2$ additional messages (control messages in \Cref{subsec:Termination}) to capture a \DTCC. We believe that it is the first such algorithm that also satisfies all other properties in \Cref{tab:ComparisonOfAlgorithms}.
    Due to the asynchronous nature, the checkpoint imposes low overhead on the transactions. Multiple checkpoints can also be executed parallelly while maintaining this property. Each parallel invocation also needs to maintain only its own set of $\ReplicaColor$ variables and message colors.
    \item \textbf{Benefits of Strict Concision and Recency:}
    \begin{enumerate}
        \item \textbf{Verification of Stable Properties:} As discussed in \Cref{subsec:DTCS}, when \DTCC is used to take multiple snapshots, it partitions the computation into \textit{super-transactions} 
        and these super-transactions provide strong consistency among them. In other words, \DTCC emulates strong consistency in an architecture that fundamentally operates on eventual consistency. This can be very useful for identifying bugs in the implementations of eventually consistent systems, which are known to be error-prone\cite{provingcorrectnessoftransformation}.
        The checkpoint also contains a precomputed state (without additional logs), enabling a quicker scan for invariant checking.
        \item \textbf{Fast disaster recovery: } Because the stored snapshot is a computed state of the database, a database can go back in time before a disaster(such as ransomware attacks, unsafe sequence of transactions by a user, database bugs, etc) almost instantly.
    \end{enumerate}

\end{enumerate}

\section{Conclusion}\label{sec:Conclusion}

In this paper, we focused on the problem of checkpointing and recovery of fully replicated database systems.
Since maintaining strong consistency often leads to high overhead, many of these systems provide weaker consistency models, such as eventual consistency.
It is well known that developing an eventually consistent data store is prone to errors \cite{provingcorrectnessoftransformation}. 
One way to assist in detecting these errors is to take frequent checkpoints to ensure that the desired invariant constraints are satisfied. 

A common problem with eventual consistency is the unpredictability caused by it. One way to address this is to periodically check if certain invariant properties are satisfied. With \DTCC, when the invariant is violated, it uniquely identifies the \textit{super-transaction} (transactions since the last snapshot) that caused this inconsistency. In this way, \DTCC mitigates the unpredictability of eventual consistency. 

To be useful, checkpointing needs to be frequent and of low overhead. Specifically, checkpoint process should be asynchronous and not block the shared memory/database operation in any way. Our checkpointing algorithm achieves this - the overhead on an operation is only to read the color variable of the database (without any locks), temporarily perform copy on writes, and to add the color (2 bits required) to messages sent by the application. 


There are hybrid main memory databases that consist of some objects that are synchronously replicated and some that are asynchronously replicated. For example, the username of a website has to be synchronously replicated to avoid duplicate usernames. However, the view count of a website can be replicated asynchronously by making the view count object a CRDT counter \cite{CRDTs}. Our algorithm can additionally be extended to these databases by taking a local snapshot consisting of both types of objects, then taking only the asynchronous update increments up to the colored cut. 

{\small
\bibliographystyle{plain}
\bibliography{ref}
}
\newpage
\appendix
\section{Complete Description of \MuFASA }\label{sec:Algorithm}

    In this section, we present our checkpointing algorithm, \MuFASA. In this algorithm, a dedicated replica, say $R1$, is the initiator that captures its local state as its local cut. Then, this local state is modified to obtain the final checkpoint that is a \DTCC. 
    First, we give an overview of \MuFASA. Then, we describe \MuFASA and prove its correctness by showing how it satisfies all of the requirements in \Cref{subsec:DTCS}. 
    For simplicity, we describe \MuFASA to take the first execution of the checkpoint. Extending it to take subsequent checkpoints with sequence numbers is discussed in \Cref{appendix:subsequent checkpoints}. The subsequent checkpoint can also take partial checkpoints that contain the incremental changes to the previous checkpoint (super transaction) by enforcing transactions to mark updated objects with a bit, as illustrated in \cite{CALC}.

    \subsection{Overview}

The key idea of the algorithm is as follows: Replica $R1$ takes a local snapshot of its own replica (using the approach from CALC \cite{CALC}) to identify its local cut. 
Then, as $R1$ learns of transactions at other replicas (which happens anyway as part of the replication process) that should be included in the checkpoint, the corresponding global checkpoint (\DTCC) is obtained. 

{A naive implementation using the color system in \cite{chandylamport,LaiYangGlobalSnapshot, MATTERN1993423, vijaygargscalableCP} would be that when $Ri, (i\neq 1)$ receives a red message, it stores its local state (database state) and becomes red.
This needs each replica to store entire copies of the database on its own, thereby, violating strict concision. 
Hence, to distinctly know the set of transactions that are not checkpointed by $R1$, but locally committed in $Ri$ before the checkpoint, we use a third color - Yellow.

Initially, all replicas are Green and send Green messages. When $Ri, (i \neq 1)$, first receives a message that is not Green, it changes to Yellow. This change must mark the local cut before any additional message passing takes place at $Ri$. $Ri$ finishes sending of all remaining update messages for transactions that were already committed when its color became yellow. Care is taken to do this concurrently without stopping the database. Then, $Ri$ changes itself to Red and sends Red messages for all transactions that were committed after its color became Yellow. Thus, Green and Yellow messages contain transactions that are meant to be a part of this checkpoint and Red messages contain transactions that are not a part of this checkpoint.

    \subsection{Elements of the Algorithm}
\MuFASA consists of 5 interrelated parts (with a sample execution in \Cref{fig:mufasaExecution}): 

\begin{figure}[t]
        \centering
        \begin{subfigure}{0.3\textwidth}
        \includegraphics[width=1.0\textwidth]{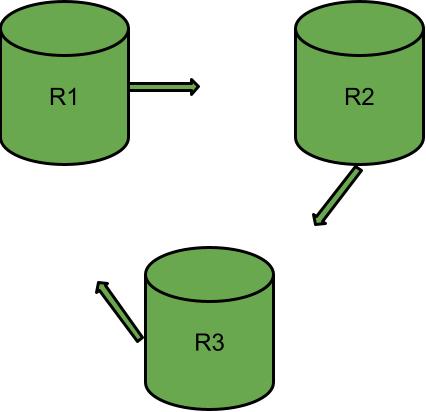}
        \caption{All replicas and their sent messages are initially Green.}
        \label{fig:Phase1}
        \end{subfigure}
        \hfill
        \begin{subfigure}{0.3\textwidth}
        \includegraphics[width=1.0\textwidth]{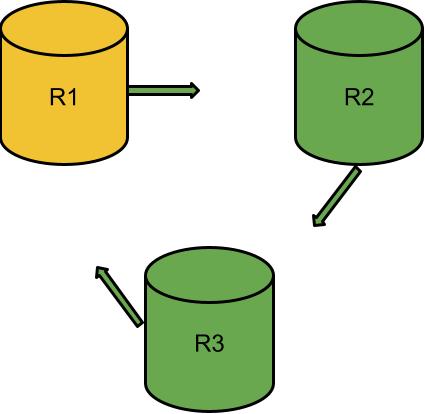}
        \caption{$R1$ turns Yellow and initiates a checkpoint CP locally.}
        \label{fig:Phase 2}
        \end{subfigure}
        \hfill
        \begin{subfigure}{0.3\textwidth}
        \includegraphics[width=1.0\textwidth]{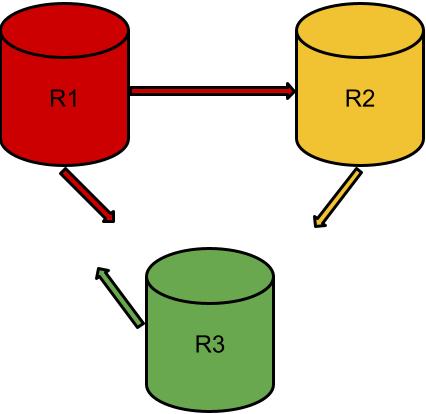}
        \caption{$R1$ turns Red and sends Red messages and $R2$ received it.}
        \label{fig:Phase 3}
        \end{subfigure}
        \hfill
        \begin{subfigure}{0.45\textwidth}
        \centering
        \includegraphics[width=0.67\textwidth]{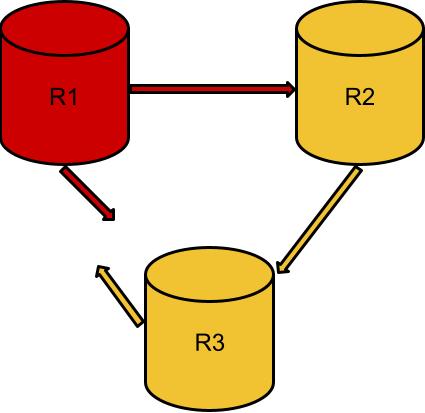}
        \caption{$R3$ asynchronously received the new color message from $R2$ first, and turns Yellow to send Yellow messages to be included in CP.}
        \label{fig:Phase 4}
        \end{subfigure}
        \hfill
        \begin{subfigure}{0.45\textwidth}
        \centering
        \includegraphics[width=0.67\textwidth]{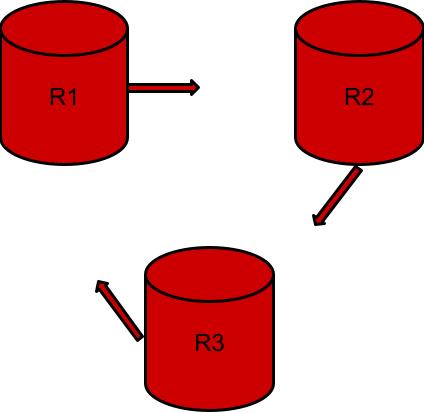}
        \caption{$R2$ and $R3$ turn Red after their $\MessageSender$ threads read $\CheckpointCutEntry$ and send Red messages that are not to be included in CP.}
        \label{fig:Phase 5}
        \end{subfigure}
        \caption{Execution of \MuFASA in a 3 replica system. Arrows (messages) that reached a replica are messages that are considered delivered, the rest are in transit.}
        \label{fig:mufasaExecution}
    \end{figure}

\begin{enumerate}[wide, labelwidth=!, labelindent=0pt, topsep=0pt]
        \item 
        \textbf{Local Database Snapshot at $R1$ for its local cut:}

        First, we utilize the approach in CALC \cite{CALC} to obtain a local database state at replica $R_1$. 
            $R1$ maintains a color variable (stores Green, Yellow or Red). Each transaction at $R1$ reads this variable at the beginning and before commit. It is only changed by the checkpointing algorithm (\Cref{Alg:AsyncCheckpointer}). Transactions that were committed when $R1$ was Green or Yellow are included in this local snapshot and the others (committed during Red) are not included in the snapshot. 
            %
            The semantics associated with the color of $R1$ is as follows:
            \begin{enumerate}[wide, labelwidth=!, labelindent=0pt]
                \item Green:  In this color, (as in \Cref{fig:Phase1}) the checkpointing thread at $R1$ waits until it gets a trigger to start a checkpoint. This trigger could be a checkpoint periodicity or something external. When the color of $R1$ is Green, all outgoing messages will have the color Green. 
                \item Yellow:  When the $\AsyncCheckpointer$ is started, color of $R1$  becomes Yellow (as in \Cref{fig:Phase 2}). 
                In this color, the checkpoint ensures that all ongoing local transactions have their start-color as Yellow. Effect of all transactions that complete with Yellow or Green contribute towards the stored checkpoint state. In this case, the outgoing messages of $R1$ will still be Green, i.e., $R1$ never sends Yellow messages. 
                \item Red Color:  $R1$ can change its color from Yellow to Red only when all transactions with Green start color have terminated (as in \Cref{fig:Phase 3}). A transaction that commits after the color is changed to Red is not included in the checkpoint.
                Hence, whenever a transaction that commits after changing the color to Red performs a write on an object for the first time, copy-on-write is used to save the previous state of the object in a stable value which is included in the checkpoint. 
                The execution point when the color is switched from Yellow to Red denotes the local cut at $R1$.
                
            \end{enumerate}
        \item \textbf{Utilizing colors to complete a global snapshot:}
        Next, we discuss how other replicas, i.e., $Ri, (i \neq 1)$ extend the local to create a \DTCC. This is achieved by associating a color with other replicas and messages sent by them. The color of messages sent by $R1$ is as discussed previously. For $Ri, (i \neq 1)$, the color of the replica and the color of the messages that it sends are the same. 
        We discuss the color of messages sent by other replicas and their interpretation to obtain the global cut next.
        
        \begin{enumerate}[wide, labelwidth=!, labelindent=0pt]
            \item Green: Initially, the replica $Ri$ is Green (as in \Cref{fig:Phase1} and $R3$ in \Cref{fig:Phase 3}) and only sends Green messages. Receiving Green messages does not trigger any checkpointing on $Ri$.
            \item Red: 
            When green $Ri$ receives a Red message (from anyone), it changes its color to Yellow (as $R2$ in \Cref{fig:Phase 3}) and appends a log entry to distinguish transactions on $Ri$ that happened before the cut (before this log entry) and those that happened after the local cut. However, changing the replica color ($\ReplicaColor$) to Yellow and the append to the log are may not be atomic.  However, we need every message sent after the cut to be Yellow. We ensure this by making every thread that interacts with the network channels ($\MessageSender$ and $\MessageReceiver$) to change the color to Yellow when it notices the color change (either through a received message or through $\ReplicaColor$) and append a log entry before proceeding to interact with the network channels. With this, if such a color change is not detected by a thread yet, its channel interaction can be said to have happened before both the color change and the cut.
                \item Yellow: A Yellow message is the same as a Red message to all Replicas $Ri, (i \neq 1)$ (If Green $Ri$ receives a Yellow message from $Rj$, it indicates that $Ri$ is learning about the checkpoint indirectly via $Rj$ like $R3$ in \Cref{fig:Phase 4}.) Hence, $Ri$ will take the same actions as for a Red message. However, Yellow has a special meaning to the initiator $R1$ - Yellow messages contain transactions that are to be included in the snapshot, as the Yellow messages are sent for transactions that had locally been completed at $Ri, (i \neq 1)$ when $Ri$ learnt of the checkpoint. These transactions have happened before the local cut at $R3$, but $R1$ was not aware of it when it created its local cut.  
                \footnote{For instance, this can happen if, say, a transaction $T3$ locally committed on $R3$ is replicated to $R2$ quickly but $R1$ slowly. Transaction $T2$ commits after receiving $T3$ in $R2$. This is quickly replicated to $R1$ (and $R3$). $R1$, receives $T2$, and then commits $T1$ without receiving $T3$. Now, if $T1$ is part of the checkpoint, then so must $T3$ . However, $R1$ is unaware of $T3$ and must wait until it receives it from $R3$.}
            \end{enumerate}
        \item \textbf{Colored Message Counters}:
            To terminate its algorithm, $R1$ needs to ensure that it has received all the Green and Yellow messages from other replicas as part of the global snapshot. To handle this issue, we utilize two counters that are inspired by \cite{MATTERN1993423, vijaygargscalableCP}, one for Green/Yellow messages and one for Red messages. This allows $R1$ to detect if all Green/Yellow messages are received from other replicas. 
        \item \textbf{Control Messages:} 
            We utilize two types of control messages in our algorithm. $\CutForCheckpoint$ is a control message sent by $R1$ to all other replicas. When replicas receive this message or the first message of a non-Green color from any replica (whichever happens first), they perform a local cut and turn themselves Yellow. Once they turn themselves Red (as described in the log entry element below), they respond to $R1$ with a $\CutForCheckpointReply(c)$ message, where c is the number of Green/Yellow messages that $R1$ must receive along that channel in its lifetime to ensure it has all necessary transactions to complete its \DTCC. 

        \item \textbf{Special Log Entries:}
            Due to asynchronous behavior between the time at which a local transaction was committed and the time at which its corresponding replication message is sent, a change in color of the replica and the local cut required is not together an atomic operation. We require color of replication messages to change to Red, during the first transaction that commits \textit{after} the cut is made at that replica. To ensure this, we add a $\commitLog$ entry, $\CheckpointCutEntry$. After performing a cut, the messages sent are Yellow until this $\CheckpointCutEntry$ log is read by the $\MessageSender$. After reading this entry, color of this replica and the color of its future replications sent by this $\MessageSender$ become Red.
    \end{enumerate}

\subsection{The Checkpointing Algorithm}
    Algorithms \ref{Alg:initialization}-\ref{Alg:AsyncCheckpointer} explain the functioning of the entire database relevant to the checkpointing algorithm. As shown in \Cref{Alg:initialization}, all replicas have their initial $\ReplicaColor$ as Green. $R1$ additionally stores $\GreenMessageCounter$ and $\RedMessageCounter$ vectors initialized with zeroes to keep track of the numbers of received messages from all replicas split by color. All other replicas maintain a $\MessageCounter$ that will contain the number of Green/Yellow messages  sent by it. \Cref{Alg:DatabaseFunctioning} describes how a database performs a transaction. Except for $R1$, all replicas perform their transactions as depicted by the system. 
    
    Replica $R1$ performs checks on the $\ReplicaColor$ to perform copy on write; the goal is to ensure that the first red transaction that performs an update on a given object creates a copy of the previous value that should be included in the checkpoint.
    Specifically, if the start color of a transaction is Green or Yellow, then a copy on write is performed, but not marked to be finalized as stable (through the $stable\_bit$ boolean). If such a transaction ends when the replica color is Red, then it marks the copied value as stable. Otherwise, either the next red transaction on the object can mark it as stable or the checkpointer marks it as stable before saving it for the checkpoint. 
    
    \Cref{Alg:AsyncCheckpointer} describes our checkpointing algorithm. First, the checkpointer changes $\ReplicaColor$ to Yellow and waits until all transactions that started during Green color have terminated.
    Then, $\ReplicaColor$ is switched to Red. We need all write-transactions that commit after this switch to not be part of the checkpoint, and the rest to be part of our checkpoint. This is ensured by the copy-on-write described in the previous paragraph.
    We note that the technique so far is the same approach as CALC \cite{CALC} to take a \textit{local} snapshot of $R1$. 
    
    An important point of interest from the CALC algorithm is that the point of switching color from Yellow to Red in $R1$ is is our desired local cut, that separates all transactions that occurred in $R1$ to have either happened before or after the checkpoint. We utilize this switch in $\MessageSender$ (as in line \ref{line:R1MsgKnowsRed}) thread of $R1$ to change the color of the messages it sends, so other replicas can help complete the checkpoint and obtain a \DTCC.
    To avoid the scenario where $R_1$ has no update message to send, we additionally send a control message, $\CutForCheckpoint$, to all replicas through the control message channels. 

    To help complete the checkpoint, all other replicas $Ri$, ($i \neq 1$) passively look for a change in color of the messages that they receive. As discussed previously, when $Ri, (i \neq 1)$ receives the first such  non-Green message, $Ri$ identifies the transactions that should be included in the checkpoint (i.e., the currently committed set of transactions in the $\commitLog$). Upon receiving the first such message at $Ri, (i \neq 1)$, $\MessageReceiver$ thread at $Ri$ switches the $\ReplicaColor$ from Green to Yellow so that future messages sent by this replica will be Yellow. 
    It additionally appends a $\CheckpointCutEntry$ into the $\commitLog$ (Line \ref{line:messagesenderlogappend} and Lines \ref{line:controlmessagelogappend} and \ref{line:messagereceiverlogappend}). This log entry marks the set of transactions that should be included in the global snapshot computed by $R1$. All subsequent ones in the $\commitLog$ occurred after the checkpoint. 
    
    Notice that the $\MessageReceiver$ threads have two instructions - (1) $\ReplicaColor$ color change from green to yellow and (2) an append to $\commitLog$ entry. This is prone to race conditions. Specifically, when a $\ReplicaColor$ change to Yellow took place without the $\commitLog$ entry being immediately added.
        \begin{wrapfigure}{L}{0.46\textwidth}
        \vspace{-0.8cm}
        \includegraphics[width=0.44\textwidth]{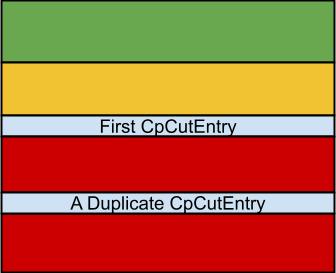}
        \caption{$\MessageSender$ interpretation\\ of the $\commitLog$ with colors.}
        \label{fig:commitLogColor}
        \vspace{-0.5cm}
    \end{wrapfigure}
    To deal with this, any $\MessageSender$ or $\MessageReceiver$ that detects that the $\ReplicaColor$ is no longer Green, ensures that it has appended a $\CheckpointCutEntry$ into the $\commitLog$ entry before proceeding with its interactions with the channels. This also ensures that the color change in $Ri$ is noticed by other replicas only after it has performed a local cut in its machine, to ensure global snapshot consistency. To ensure both of these steps are both non-blocking, all of these threads use a compare and swap operation to conditionally change $\ReplicaColor$ from Green to Yellow (only $\MessageSender$ can change it to Red as will be explained next). 
    $\MessageSender$ will act on the first $\CheckpointCutEntry$ to change $\ReplicaColor$ and its subsequent sent messages to Red, and ignore the other duplicate entries to ensure that the subsequent appends of this entry do not affect the checkpointing technique. 

    $\MessageReceiver$ must additionally ensure no replicated transactions get executed before $\CheckpointCutEntry$ is appended. For this, it additionally has a thread local variable called appendSuccess, which will be set to true after the append. When a future $\MessageReceiver$ thread receives a Yellow or Red message for the first time, and the $\ReplicaColor$ is still Yellow, it appends a log if appendSuccess is false. This way each $\MessageReceiver$ thread helps complete the checkpoint by appending $\CheckpointCutEntry$ to the log atmost once.

    After $\MessageSender$ reads the $\CheckpointCutEntry$ log entry, each replica $Ri, (i \neq 1)$ sends a $\CutForCheckpointReply$ with its $\MessageCounter$ value. This is utilized by $R1$ to know when the checkpointer (\Cref{Alg:AsyncCheckpointer}) can terminate. As described in line \ref{line:R1copyingremotetxn} $R1$ additionally applies its received Green/Yellow transactions into the checkpoint that it just captured, to ensure it is a \DTCC. The checkpointer finally terminates when all of the $\GreenMessageCounter$ values match the $\MessageCounter$ values sent by its remote replicas.

\begin{algorithm}[!h]
    \caption{Initialization of the Database.}
    \label{Alg:initialization}
    \scriptsize
    \DontPrintSemicolon
    \textbf{Common initialization of state at all Replicas $\neq R1$:}
    
    \Indp Integer $\ReplicaColor$ $\gets$ GREEN \;
    Integer $\MessageCounter$ $\gets$ $0$\;
    
    \Indm \textbf{Additional initial state variables at $R1$:}

    \Indp Integer $\ReplicaColor$ $\gets$ GREEN \;
    Integer $\GreenMessageCounter[2\cdots n]$ $\gets$ $[0, \cdots, 0]$\;
    Integer $\RedMessageCounter[2 \cdots n]$ $\gets$ $[0, \cdots, 0]$\;
    
    
    
    bit stable\_status[DB\_SIZE] $\gets$ $[not\_available, \cdots, not\_available]$ \;
    
    \ForEach{\normalfont key in DB}{
        db[key].live $\gets$ initial value \Comment{\normalfont Replicas $\neq R1$ will only have live values}
        
        db[key].stable $\gets$ $\phi$\;
    }

\vspace*{-5mm}
\end{algorithm}

\begin{algorithm}[!h]
    \caption{Functioning of the Database.}
    \label{Alg:DatabaseFunctioning}
    \scriptsize
    \DontPrintSemicolon
    \begin{multicols}{2}
    \Fn{\normalfont\textbf{ApplyWrite}{(txn, key, value)} \Comment*[h]{At Replica $R1$}} {
        \eIf{\normalfont txn.start-color $=$ YELLOW} {
            \If{\normalfont stable\_status[key] $=$ not\_available} {
            db[key].stable $\gets$ db[key].live\;
            }
        } {
            \eIf{\normalfont txn.start-color $=$ RED} {
                \If{\normalfont stable\_status[key] $=$ not\_available} {
                db[key].stable $\gets$ db[key].live\;
                stable\_status[key] $\gets$ available\;
                }
            } {
                \If{\normalfont txn.start-color $=$ GREEN} {
                    \If{\normalfont db[key].stable $\neq$ $\phi$} {
                    Erase db[key].stable\;
                    }
                }
            }
        }
        db[key].live $\gets$ value\;
    }


    \Fn{\normalfont\textbf{Execute}(txn) \Comment*[h] {\normalfont At Replica $R1$}} {
        txn.start-color $\gets$ $\ReplicaColor$\;
        Request txn's locks\;
        Run txn logic, using \textbf{ApplyWrite} for updates\;
        txn.end-color $\gets$ $\ReplicaColor$\;

        Append txn commit token to $\commitLog$\;
        \If{\normalfont txn.start-color $=$ YELLOW} {
            \eIf{\normalfont txn.end-color $=$ YELLOW} {
                \ForEach{\normalfont foreach key in txn} {
                
                    Erase db[key].stable\;
                }
            } {
                \If{\normalfont txn.end-color $=$ RED} {
                    \ForEach{\normalfont foreach key in txn} {
                    
                        stable\_status[key] $\gets$ available\;
                    }
                }
            }
        }
        Release txn's locks\;
    }

    \Fn{\normalfont\textbf{Execute}(txn) \Comment*[h] {\normalfont At Replica $\neq R1$}} {
        Request txn's locks\;
        Run txn logic for updates\;
        Append txn commit token to $\commitLog$\;
        Release txn's locks\;
    }
    \end{multicols}
\vspace*{-5mm}
\end{algorithm}

\begin{algorithm}[!h]
    \caption{$\AsyncCheckpointer$ and Message Processing for \MuFASA.}
    \label{Alg:AsyncCheckpointer}
    \scriptsize
    \DontPrintSemicolon
    \setlength\columnsep{12pt}
    \begin{multicols}{2}
    \Fn{\normalfont $\AsyncCheckpointer$ at Replica $R1$} {
        $\ReplicaColor$ $\gets$ YELLOW \;
        wait until all active txns have start-color YELLOW\;

        $\ReplicaColor$ $\gets$ RED \;\label{line:R1changetored}

        wait until all active txns to have start-color RED\;

        \ForEach{\normalfont key in db} {
        
            \eIf{stable\_status[key] $=$ available} {
                Write db[key].stable to checkpoint\;
                Erase db[key].stable\;
            } {
                \If {stable\_status[key] = not\_available} {
                    stable\_status[key] $\gets$ available\;
                    val $\gets$ db[key].live\;
                    \eIf{db[key].stable $\neq \phi$ } {
                    Write db[key].stable to checkpoint\;
                    Erase db[key].stable\;
                    } {
                        \If{db[key].stable $= \phi$} {
                        Write val to checkpoint\;
                        }
                        
                    }
                    
                } 
            }
        }

        Send $\CutForCheckpoint$ to all replicas $\neq R1$ through control message channels and wait for $\CutForCheckpointReply$$[2\cdots n]$;\label{line:SendcontrolMsgFromR1}

            wait for $\GreenMessageCounter$$[2\cdots n]$ to match the counters returned in $\CutForCheckpointReply$$[2\cdots n]$ through the control message channels\; \label{line:WaitForCounterMatch}

    }

    \Fn{\normalfont $\MessageSender$(M) for Replica $R1$} {
        \While{\normalfont true} {
            entry $\gets$ the next entry in the $\commitLog$\ \Comment{\normalfont Waits for new entries}

            \If{\normalfont entry is a local transaction}{
                msg $\gets$ create a replicate message corresponding to txn in the entry\; 
                          
                \eIf{\normalfont $\ReplicaColor$ $=$ RED \label{line:R1MsgKnowsRed}} {
                    msg.color $\gets$ RED\;
                } {
                    msg.color $\gets$ GREEN\;
                }
                Add msg to all channel send queues\;                
            }
        } 
    }
    
        \Fn{\normalfont $\MessageReceiver_{me}^j$(M) for $R1$} {
        $txn \gets \ExecuteTransaction(M.txn)$\;
        \If{\normalfont txn.end-color $=$ RED $\land$ M.color $\neq$ RED\label{line:R1CheckingMsgColor} } {
            Use $\ExecuteTransaction$(M.txn) on the checkpoint\;\label{line:R1copyingremotetxn}
        }
        \eIf{\normalfont M.color $=$ GREEN $\lor$ M.color $=$ YELLOW} {
            $\GreenMessageCounter$[j]++\;   \label{line:GreenCountIncrementR1}
        } {
            $\RedMessageCounter$[j]++\;
        }
    }

    \Fn{\normalfont upon receiving $\CheckpointCutEntry$ from $R1$ } {
            CAS($\ReplicaColor$, GREEN, YELLOW)\;\label{line:CASControlMessage}
            Append $\CheckpointCutEntry$ into $\commitLog$\;\label{line:controlmessagelogappend}
    }
    
    \Fn{\normalfont $\MessageReceiver_{me}^j$(M) for any Replica $\neq R1$} {            
        \eIf{\normalfont $\ReplicaColor$=GREEN $\land$ $M.color$$\neq$GREEN } {
            CAS($\ReplicaColor$, GREEN, YELLOW)\;\label{line:CASMessageReceiver}
           Append $\CheckpointCutEntry$ into $\commitLog$\;\label{line:messagereceiverlogappend}
           appendSuccess $\gets$ true\;
       }{
       \If{\normalfont $\ReplicaColor$=YELLOW $\land$ appendSuccess=false}{
            Append $\CheckpointCutEntry$ into $\commitLog$\;\label{line:messagereceiverlogappendOnFalse}
            appendSuccess $\gets$ true\;
       }
       }
        $\ExecuteTransaction$(M.txn)\;\label{line:remoteTxnAtRi}
    }


    \Fn{\normalfont $\MessageSender$(M) for any Replica $\neq R1$} {
        $\currentColor$ $\gets$ GREEN\;
        \While{\normalfont true} {
            entry $\gets$ the next entry in the $\commitLog$\ \Comment{\normalfont Waits for new entries}
            \eIf{\normalfont entry is a local transaction}{
                msg $\gets$ create a replicate message corresponding to txn in the entry\; 
                \If{\normalfont $\currentColor$ $\neq$ $\ReplicaColor$} {
                    \If{\normalfont $\ReplicaColor$ $=$ YELLOW} {
                        Append $\CheckpointCutEntry$ to $\commitLog$\;\label{line:messagesenderlogappend}
                        $\currentColor \gets \ReplicaColor$\; \label{line:currentColorBecomesYellow}
                    }
                }
                msg.color $\gets$ $\currentColor$\;\label{line:MessageColorSetRi}
                $\MessageCounter$++\;\label{line:RiMessageCounterInc}
                
                Add msg to all channel send queues\;                
            } {
                \If{\normalfont entry is of type $\CheckpointCutEntry$ \label{line:MessageSenderCpEntryCheck}} {
                    \If{$\ReplicaColor$ $\neq$ RED} {
                        $\ReplicaColor$ $\gets$ RED\; \label{Line:RiTurnsRed}
                        $\currentColor$ $\gets$ $\ReplicaColor$\;\label{line:currentColorBecomesRed}
                        Send $\CutForCheckpointReply(\MessageCounter$) to $R1$ through the control message channel\;
                        $\MessageCounter$ $\gets$ 0\;
                    }
                }
            }
        }
    }
    \end{multicols}

\vspace*{-5mm}
\end{algorithm}

\section{Subsequent checkpoints}\label{appendix:subsequent checkpoints}
    In this section, we explain how subsequent checkpoints can be taken after the first checkpoint is completed. For this, instead of attaching a color to every replicate message, we attach a checkpoint number ($cpNum$) to every message. The number starts from 1. The starting number $cpNum$ indicates Green color of the messages, $cpNum+1$ indicates Yellow color and $cpNum+2$ indicates Red color. For the first checkpoint, we utilize numbers 1, 2 and 3 to stand for colors Green, Yellow and Red. Note that, at the end of the first checkpoint, all messages will contain the checkpoint number 3. Thus, for the second checkpoint, we start from 3, to stand for Green (formerly Red), 4 for Yellow and 5 for Red. The third checkpoint will alternate 5 to mean Green and so on. All updates to color are instead seen as updates to $cpNum$.
    
    To avoid the problem of duplicate logs, we distinguish the $\CheckpointCutEntry$ logs using the current $cpNum$. Upon detecting a $\CheckpointCutEntry$ of a specific $cpNum$, the $\MessageSender$ skips all subsequent $\CheckpointCutEntry$ of any smaller $cpNum$. This ensures that duplicates are not processed and that the system responds only to new checkpoint initiations. The same goes for color change detection in the $\MessageReceiver$s. When the replica is green (as detected by $cpNum$) only messages with a different $cpNum$, which is of value greater than the replica's current $cpNum$ is used for detecting this change in color. This way, the system can perform checkpoints one after another. When the checkpointer waits for the message counter values, it chooses between Red and Green counters depending on the current checkpoint number. if $cpNum$ is 1 modulo 4, then it uses $\GreenMessageCounter$, and if $cpNum$ is 3 modulo 4, it uses $\RedMessageCounter$. After waiting for these counters, they will not be incremented before a subsequent checkpoint. So we can safely reset the currently used counter array to 0.  As for other variables, instead of resetting $stable\_status$ to $not\_available$, we instead swap $not_available$ and $available$ values because in one iteration of the checkpoint $stable\_status$ bits of all objects are flipped, so it is faster to change the meaning of 0s and 1s. This bit flip technique is borrowed from CALC\cite{CALC}.  

    It is also worth noting that the $cpNum$ counter can be bounded by rotating checkpointing initiation responsibility to different replicas in a round robin format.

\section{Correctness of \MuFASA}\label{appendix:correctness}
    To show the correctness of \MuFASA, we need to show that the final checkpoint captured by \MuFASA is \DTCC and that the \MuFASA algorithm satisfies all 4 desired properties listed in \Cref{subsec:DTCS}}. Wherever required, we borrow the guarantees of CALC \cite{CALC}, without proving their results again in this paper.
    We first make the following observations about the algorithm:
    \begin{observation}\label{obs:FinalCPSequence}
        The final checkpoint stored by \MuFASA is a snapshot of all objects in $R1$ using CALC, followed by a series of Green and Yellow message remote transactions executed on top of this snapshot. 
    \end{observation}
    \begin{proof}
        This is observable from the fact that the only place where the checkpoint is modified after CALC algorithm is at \Cref{line:R1copyingremotetxn}. The message color of the remote transaction is always checked to not be Red before they are applied. 
    \end{proof}
    \begin{observation}\label{obs:finiteSteps}
        The \MuFASA algorithm only adds a finite number of additional steps to the transaction and message passing frameworks of the distributed database.
    \end{observation}
    \begin{proof}
        By finite, here we mean that there is no indefinite wait on the checkpointing thread.
        For the database transactions, additional instructions are present in $\tt{Execute}$, $\tt{ApplyWrite}$ methods of Replica $R1$ (refer to \Cref{Alg:DatabaseFunctioning}). Th
        The total number of additional steps here are in the order of the number of keys in the transaction, as it is a fixed number of extra instructions in a loop over the keys used in the transaction. Replica $Ri, i \neq 1$ takes no additional steps. For the message passing framework, the overhead added to the processing of the message in $\MessageSender$ and $\MessageReceiver$ of any Replica is constant (refer to \Cref{Alg:AsyncCheckpointer}). Hence \MuFASA only adds a finite number of total steps to any aspect of the distributed database. 
    \end{proof}
    \begin{observation}\label{obs:RiGreen}
        $\ReplicaColor$ at $Ri$ is only set to green during initialization and never set to Green again in the future.
    \end{observation}
    \begin{proof}
        This observation is trivial, as there are only $\ReplicaColor$ assignments at $Ri$ for Yellow or Red. $\ReplicaColor$ is only initialized with Green in \Cref{Alg:initialization}.  
    \end{proof}
    \begin{observation}\label{obs:RiTurnsYellow}
        $\ReplicaColor$ at $Ri, i \neq 1$ changes to Yellow only once and it is before the delivery(processing) of the first non-green message received at $Ri$.
    \end{observation}
    \begin{proof}
        $\ReplicaColor$ at $Ri$ is only ever set to Yellow by the CAS operations at \Cref{line:CASControlMessage} and \Cref{line:CASMessageReceiver}. Consider multiple $\MessageReceiver$ threads simultaneously receiving messages of color Yellow or Red when $\ReplicaColor$ is still Green. Due to \Cref{obs:RiGreen}, only one of these threads can succeed in the compare and swap operation to set it to Yellow. Future Yellow or Red $\MessageReceiver$ threads will not invoke this operation, as $\ReplicaColor$ is no longer Green.  
    \end{proof}
        \begin{observation}\label{obs:RiRed}
        The $\MessageSender$ at $Ri, i \neq 1$ sets $\ReplicaColor$ to Red only once and it is after it reads the first $\CheckpointCutEntry$ $\commitLog$ entry.
    \end{observation}

    \begin{proof}
        This is observable from the fact that Red is assigned to $\ReplicaColor$ at $Ri$ only at \Cref{Line:RiTurnsRed}, and only by the $\MessageSender$ thread. Thus, the condition before it that checks for Red ensures that this takes place only when the $\MessageSender$ thread reads a $\CheckpointCutEntry$ entry for the first time in \Cref{line:MessageSenderCpEntryCheck}.  
    \end{proof}
    \begin{observation}\label{obs:Only3Colors}
        The $\ReplicaColor$ variable can only take 3 values - Green, Yellow and Red at any replica $Ri, i \neq 1$.
    \end{observation}
    \begin{proof}
        This is trivial, as the initial value of $\ReplicaColor$ is Green and any modification to it are by  \Cref{line:CASControlMessage} and \Cref{line:CASMessageReceiver} setting it to Red and \Cref{line:R1changetored} that sets it to Red. Hence Green, Yellow and Red are the only possible values for $\ReplicaColor$.
    \end{proof}
    \begin{observation}\label{obs:CutEntryAfterYellow}
        The first $\CheckpointCutEntry$ log entry appended to the $\commitLog$ at $Ri, i \neq 1$ when the $\ReplicaColor$ is  Yellow.
    \end{observation}
    \begin{proof}
        The three possibilities for $\ReplicaColor$ are Green, Yellow and Red by \Cref{obs:Only3Colors}. 
        $\CheckpointCutEntry$ is logged by \Cref{line:controlmessagelogappend}, \Cref{line:messagereceiverlogappend}, \Cref{line:messagereceiverlogappendOnFalse} and \Cref{line:messagesenderlogappend}. All four lines require the current color to be Yellow a priori either by a read or by a CAS. Since a color change from Green to Yellow has happened before this log append, and Green is never set to be $\ReplicaColor$ (from \Cref{obs:RiGreen}), $\ReplicaColor$ cannot be Green.
        
        \Cref{obs:RiRed} guarantees that the first $\CheckpointCutEntry$ was already present when $\ReplicaColor$ is set to Red. Hence the $\ReplicaColor$ cannot be Red.

        Thus, by the process of elimination, the $\ReplicaColor$ must have been Yellow when the first $\CheckpointCutEntry$ was appended.
    \end{proof}

    We now have the following lemma that is crucial for our correctness:
    \begin{lemma}\label{lemma:CutEntryBeforeMessage}
        The first $\CheckpointCutEntry$ log entry added to the $\commitLog$ at $Ri, i \neq 1$ takes place before the delivery(complete processing) of the first non-Green message received at $Ri$.
    \end{lemma}
    \begin{proof}
        $\CheckpointCutEntry$ is logged by \Cref{line:controlmessagelogappend}, \Cref{line:messagereceiverlogappend}, \Cref{line:messagereceiverlogappendOnFalse} and \Cref{line:messagesenderlogappend}. The instant after $\ReplicaColor$ is set to Yellow by the CAS winner of \Cref{line:CASControlMessage} and \Cref{line:CASMessageReceiver}, all threads have this new color visible due to the linearizability of the $\ReplicaColor$ variable. Thus, every remote transaction executed by \Cref{line:remoteTxnAtRi} beyond this linearization point would have the following cases in the message checks of the $\MessageReceiver$ call to it:
        \begin{enumerate}
            \item $\ReplicaColor$ was Green: Then \Cref{line:messagereceiverlogappend} would have taken place before the transaction at \Cref{line:remoteTxnAtRi}.
            \item $\ReplicaColor$ was Yellow: Then \Cref{line:messagereceiverlogappendOnFalse} would have taken place before the transaction at \Cref{line:remoteTxnAtRi}.
            \item $\ReplicaColor$ was Red: Then $\MessageSender$ has already read the first $\CheckpointCutEntry$ from $\commitLog$ before setting it to red by \Cref{obs:RiRed}. Hence, it was appended before the transaction at \Cref{line:remoteTxnAtRi}.
        \end{enumerate} 
        In any case, the transaction of first non-Green message is not executed until $\CheckpointCutEntry$ is appended into the $\commitLog$. In other words, the message has not been delivered before appending the $\CheckpointCutEntry$. 
    \end{proof}

    We also have some additional observations:
    \begin{observation}\label{obs:redColorThenRedMessage}
        After $\ReplicaColor$ at $Ri, i\neq 1$ becomes Red, the $\MessageSender$ thread of $Ri$ only sends Red messages. All other messages are not Red.
    \end{observation}
    \begin{proof}
        When $\ReplicaColor$ is set to Red in \Cref{Line:RiTurnsRed}, it is succeeded by \Cref{line:currentColorBecomesRed} that sets $\currentColor$ to Red. This $\currentColor$ is used by any future message sent by $Ri$ in \Cref{line:MessageColorSetRi}. Until then, $\currentColor$ is never set to Red, and hence messages sent before this point are not Red.
    \end{proof}
    \begin{observation}\label{obs:totalReceivedAtR1}
        The $\AsyncCheckpointer$ receives the exact number of Green and Yellow messages sent by each replica $Ri, i \neq 1$ in its lifetime.
    \end{observation}
    \begin{proof}
         The $\MessageSender$ at $Ri$ maintains a $\MessageCounter$ that is incremented upon sending messages in \Cref{line:RiMessageCounterInc}. When $\ReplicaColor$ is set to red, its future messages and only its future messages are Red (\Cref{obs:redColorThenRedMessage}). Hence, $\MessageCounter$ value at this point is the total number of Green and Yellow messages sent by $Ri$. This value is sent back to $R1$ as $\CutForCheckpointReply$ message.
    \end{proof}
    \begin{observation}\label{obs:termination}
        The $\AsyncCheckpointer$ terminates only after all Green and Yellow remote transactions from all replicas get committed at $R1$. 
    \end{observation}
    \begin{proof}
         \Cref{line:WaitForCounterMatch} terminates after the $\GreenMessageCounter$ index for each replica match with the total they had sent in their $\CutForCheckpointReply$. Since the $\GreenMessageCounter$ is only incremented in $R1$ at \Cref{line:GreenCountIncrementR1} after executing the remote transaction on the checkpoint at \Cref{line:R1copyingremotetxn}, $R1$ terminates only after all Green and Yellow remote transactions from all replicas get committed at $R1$.
    \end{proof}

We now prove the following lemma that will be useful for claiming Dependency Closure.
    \begin{lemma}\label{lemma:NoMissedTransactions}
        If there is a transaction T that is included in the final checkpoint and a transaction U has committed in a replica $Ri$ before transaction T in the issued replica $Ri$ of T, then transaction U is also included in the final checkpoint.  
    \end{lemma}
    \begin{proof}
        We analyze this scenario with all possible cases:
        \begin{enumerate}
            \item $i=1$:  CALC\cite{CALC} already guarantees that its  checkpointing algorithm is transaction consistent. Hence U will also be included in the final checkpoint by CALC.
            \item $i \neq 1$: If CALC does not already capture U, we have the following three sub cases:
            \begin{enumerate}
                \item U was issued at $Ri$: For T to be included in the checkpoint, this in turn has three subcases:
                \begin{enumerate}
                    \item Both T and U were captured by CALC: This is self-explanatory.
                    \item T was captured by CALC when U was not replicated to $R1$ yet: This means both T and U were sent in green color by the $\MessageSender$ of $Ri$. Hence U will eventually be included in the final checkpoint.
                    \item T was replicated as a Yellow or Green message and U was not captured by CALC: Since U is present before T in the $\commitLog$ of $Ri$, U will also be colored Green or Yellow. Hence U will be captured in the final checkpoint. 
                \end{enumerate}
                \item U was issued at $Rj, j \notin \{1,i\}$: For T to be included in the checkpoint, we have the same subcases like before:
                \begin{enumerate}
                    \item Both T and U were captured by CALC: This is self-explanatory.
                    \item T was captured by CALC when U was not replicated to $R1$ yet: However, we know that U has been replicated to $Ri$ already. Hence, the color of the message was decided before CALC was completed. Thus, U was replicated as a green message and will be included in the final checkpoint.
                    \item T was replicated as a Yellow or Green message and U was not captured by CALC: In this case U must have not been replicated as a Red message by $Ri$. We show this by contradiction. Let us say $Rj$ replicated U as a Red message to $Ri$. When $Ri$ receives this Red update message, it would have already inserted $\CheckpointCutEntry$ (\Cref{lemma:CutEntryBeforeMessage}) before committing U at $Ri$. Thus, any transaction after U, will not be part of the final checkpoint. Thus transaction T, which was issued in $Ri$ after U was committed in $Ri$ will also not be included in the final checkpoint.    
                \end{enumerate}
                \item U was issued at $R1$: This is impossible and is reasoned in away similar to the previous impossible case. If T is included in the checkpoint and U is not a part of CALC, then by the properties of the CALC algorithm, U happened after CALC. Hence replication of U was done with a Red message. $Ri$ that receives U would have already appended $\CheckpointCutEntry$ (\Cref{lemma:CutEntryBeforeMessage}) before performing transaction U on it. Hence the final checkpoint does not include transactions that were issued in $Ri$ after transaction U is committed in $Ri$. Hence T could have not been part of the final checkpoint.  
            \end{enumerate} 
        \end{enumerate}
    \end{proof}
    \subsection{\MuFASA records a \DTCC}
    From \Cref{subsec:DTCS}, a \DTCC must satisfy the Dirty Write Exclusion and Dependency Closure properties:
\begin{itemize}
    \item \textbf{Dirty Write Exclusion:} From \Cref{obs:FinalCPSequence}, the captured checkpoint is a checkpoint generated by CALC\cite{CALC}, that is then augmented by remote transactions executed in \Cref{line:R1copyingremotetxn}. CALC work \cite{CALC}  guarantees that its checkpoint is transaction consistent and hence has no dirty writes in its checkpoint. From \Cref{obs:termination}, \MuFASA only terminates after the remote transactions executed in \Cref{line:R1copyingremotetxn} commit. Thus, the checkpoint recorded by \MuFASA consists only of committed transactions. Moreover, the $CP_{set}$ consists exactly of the transactions recorded by CALC, and the Green/Yellow remote transactions from \Cref{line:R1copyingremotetxn}.
    \item \textbf{Dependency Closure:} This property is precisely captured by \Cref{lemma:NoMissedTransactions}. 
\end{itemize}

\subsection{Properties of \MuFASA}
The desired properties in \Cref{subsec:DTCS} are also satisfied as follows:
\begin{itemize}
    \item \textbf{Recency: } We borrow a result from CALC\cite{CALC} that the CALC technique used at R1 ensures that all transactions that committed before the $\AsyncCheckpointer$ was invoked are all included in the final checkpoint. By Dependency Closure Property of \DTCC, this means that all transactions that took place in any replica, before the invocation of the $\AsyncCheckpointer$, are also included in the final checkpoint. Thus, \MuFASA also guarantees recency of the stored snapshot.
    \item \textbf{Strict Concision:} This follows from how we initially capture the database state at $R1$, then only apply the additional remote transactions on this captured state to reach a final database state (\Cref{obs:FinalCPSequence}).
    \item \textbf{Bounded Message Complexity:} For $n$ replicas, the only additional messages exchanged are the $n-1$ $\CutForCheckpoint$ control messages from $R1$ and their $n-1$ replies (line \ref{line:SendcontrolMsgFromR1}). Hence, the  number of control messages is bounded to be $2n-2$, and does not depend on the number of objects in the database.
    \item \textbf{Asynchronous:} Since we only add finite number of additional steps to the database framework (from \Cref{obs:finiteSteps}),  the checkpointing thread never blocks the progress of transactions. Additionally, since the control messages sent are through a separate channel to avoid congestion with replication update messages, the algorithm is fully asynchronous.
\end{itemize}
\end{document}